\documentclass[%
aip,
sd,%
amsmath,amssymb,
author-numerical,%
]{revtex4-1}

\usepackage{parskip}
\usepackage{lineno}

\usepackage{amsmath}
\usepackage{amssymb}
\usepackage{amsfonts}
\usepackage{mathtools}
\usepackage{mathptmx}
\usepackage{bm}
\usepackage{lipsum}

\usepackage{array}
\usepackage{enumitem}
\usepackage{tabularx}
\usepackage{enumitem}
\usepackage{multirow}
\usepackage{dcolumn}

\usepackage{times}
\usepackage{anyfontsize}

\usepackage[pdftex]{graphicx}
\graphicspath{{./figures/}}
\usepackage{float}
\usepackage{caption}
\usepackage{float}
\usepackage{tikz}
\usetikzlibrary{arrows}
\usetikzlibrary{calc}
\usepackage{subfig}
\usepackage{pgfplots}
\pgfplotsset{compat=1.7}
\usetikzlibrary{shapes}
\usepackage{rotating}

\usepackage{url}
\usepackage{color}
\usepackage{csquotes}
\usepackage{lipsum}
\usepackage{standalone}

\graphicspath{ {home/downloads/} }

\begin{document}

	\title[]{Optimal Control of Networks in the presence of Attackers and Defenders}
	
	\author{Ishan Kafle}%
	\email{ikafle@unm.edu.}
	\affiliation{Department of Mechanical Engineering, University of New Mexico, Albuquerque, New Mexico 87131, USA}%
	\author{Sudarshan Bartaula}%
	\email{sbartaula@unm.edu.}
	\affiliation{Department of Mechanical Engineering, University of New Mexico, Albuquerque, New Mexico 87131, USA}%
	\author{Afroza Shirin}%
	\email{ashirin@unm.edu.}
	\affiliation{Department of Mechanical Engineering, University of New Mexico, Albuquerque, New Mexico 87131, USA}%
	\author{Isaac Klickstein}%
	\email{iklick@unm.edu.}
	\affiliation{Department of Mechanical Engineering, University of New Mexico, Albuquerque, New Mexico 87131, USA}%
	\author{Pankaz Das}%
	\email{pankazdas@unm.edu.}
	\affiliation{Department of Electrical and Computer Engineering, University of New Mexico, Albuquerque, New Mexico 87131, USA}%
	\author{Francesco Sorrentino}%
	\email{fsorrent@unm.edu.}
	\affiliation{Department of Mechanical Engineering, University of New Mexico, Albuquerque, New Mexico 87131, USA}%
	
	\date{\today}
	
	\begin{abstract}
		We consider the problem of a dynamical network whose dynamics is subject to external perturbations (`attacks') locally applied at a subset of the network nodes. We assume that the network has an ability to defend itself against attacks with appropriate countermeasures, which we model as actuators located at (another) subset of the network nodes. We derive the optimal defense strategy as an optimal control problem. We see that the network topology, as well as the distribution of \emph{attackers} and \emph{defenders} over the network affect the optimal control solution and the minimum control energy.  We study the optimal control defense strategy for several network topologies, including chain networks, star networks, ring networks, and scale free networks.
	\end{abstract}
	
	\maketitle
	
	\begin{quotation}
		
		Optimal control of networks is an area of recent interest in the literature, where focus has been placed on how the network topology and the position of \emph{driver} and \emph{target} nodes affect the optimal solution. Here we study a different but related problem, that of optimally controlling a network under attack. We investigate the role of the network topology as well as of the distribution of \emph{attackers} and \emph{defenders} over the network.
		Some of our results are counter-intuitive, as we find that for small chain networks, star networks, and ring networks, the distance between a single attacker and a single defender is not the key factor that determines the minimum control energy. We also consider the case of a large scale-free network in the presence of a single attacker and multiple defenders for which we see that the minimum control energy varies over different orders of magnitude as the position of the attacker is changed over the network. For this case we observe that the minimum distance between the  attacker node and the defender nodes is a good predictor of the \emph{strength} of an attack. 
	\end{quotation}

	\newcommand{\markerone}{\raisebox{0.5pt}{\tikz{\node[draw,scale=0.4,circle,fill=black!20!blue](){};}}}
	\newcommand{\markerfour}{\raisebox{0.5pt}
		{\tikz{\node[draw,scale=0.4,regular polygon, regular polygon sides=4,fill=red](){};}}}
	\newcommand{\markertwo}{\raisebox{0.5pt}{\tikz{\node[draw,scale=0.4,circle,fill=black!20!red](){};}}}
	\newcommand{\markerthree}{\raisebox{0.5pt}{\tikz{\node[draw,scale=0.4,circle,fill=black!20!green](){};}}}
	\newcommand{\markerfive}{\raisebox{0.5pt}
		{\tikz{\node[draw,scale=0.4,regular polygon, regular polygon sides=4,fill=red](){};}}}
	\newcommand{\markersix}{\raisebox{0.5pt}
		{\tikz{\node[draw,scale=0.4,regular polygon, regular polygon sides=4,fill=green](){};}}}
	\newcommand{\markerseven}{\raisebox{0.5pt}
		{\tikz{\node[draw,scale=0.4,regular polygon, regular polygon sides=4,fill=blue](){};}}}
	
	\maketitle
	
	\section{Model}
	
\begin{figure}[H]
		\centering
			\begin{tabular}{lcr}
				\begin{tabular}{c}
					\includegraphics{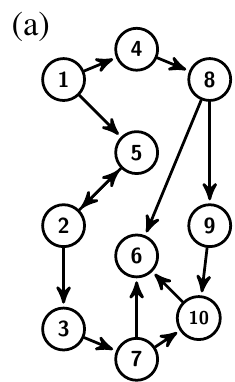}\\[9ex]
				\end{tabular}& &
				\begin{tabular}{c}
					\includegraphics[width=9cm]{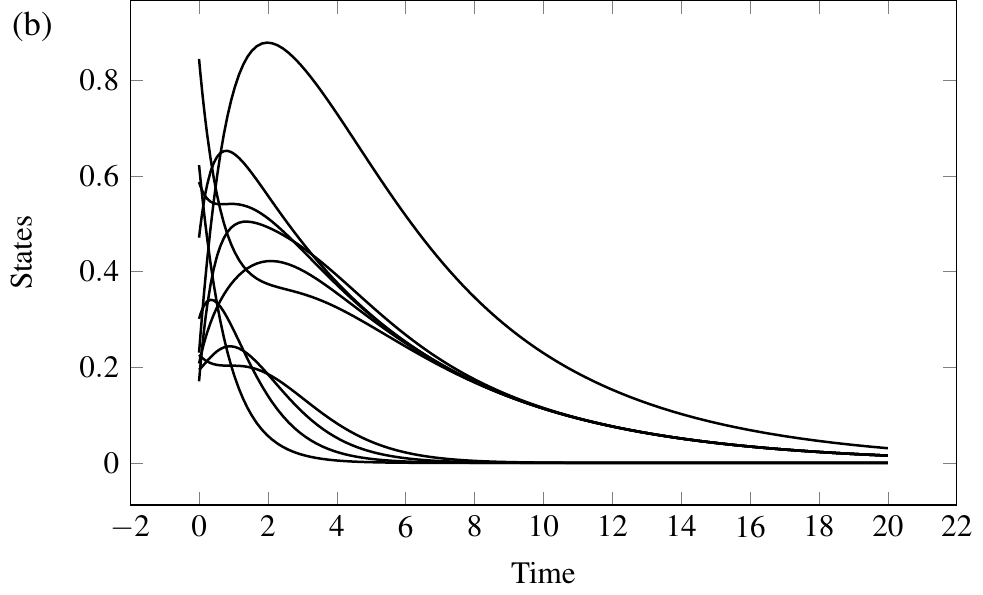}
				\end{tabular}\\
				
				\begin{tabular}{c}
					\includegraphics{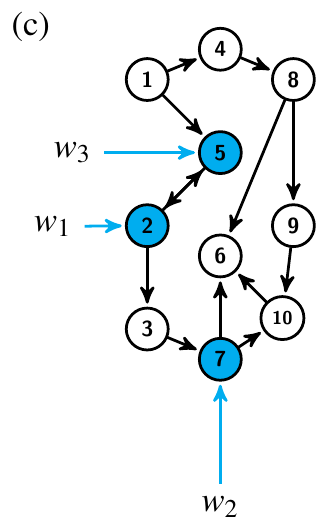}\\[10ex]
				\end{tabular}& &
				\begin{tabular}{c}
					\includegraphics[width=9cm]{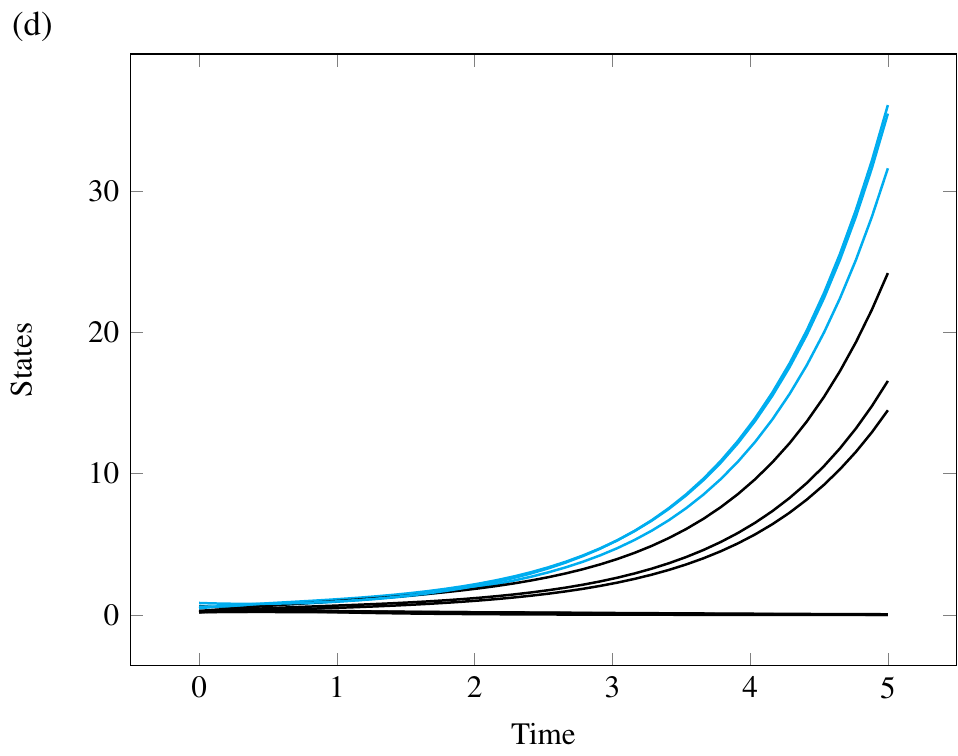}
				\end{tabular}\\
				
				\begin{tabular}{c}
					\includegraphics{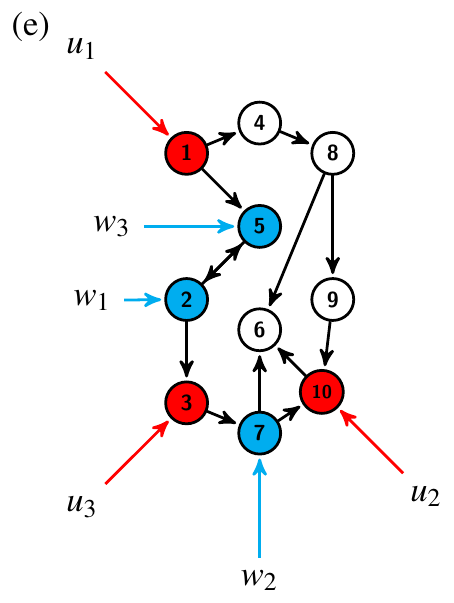}\\[5ex]
				\end{tabular}& &
				\begin{tabular}{c}
					\includegraphics[width=9cm]{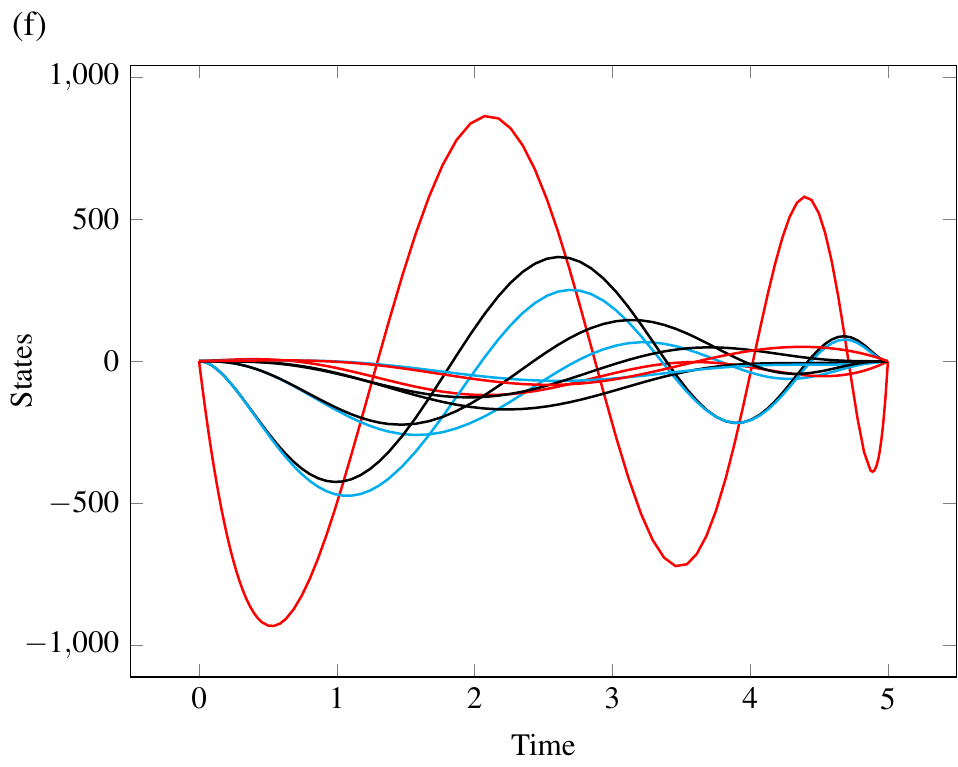}
				\end{tabular}
			\end{tabular}
		
			\caption{(a) Network under no attack. (b) Time evolution of the network nodes under no attack. (c) Same network as in (a), with attackers located at nodes 2, 5, and 7. (d) Time evolution of the network nodes under attack. (e) Same network as in (c) with defenders located at nodes 1, 3, and 10. (f) Time evolution of the network nodes under attack and response of the defender nodes.}
			\label{fig:network1}
	\end{figure}

	Most infrastructure systems are networked by design, such as power grids \cite{pagani2013power}, road systems \cite{yang1998models}, telecommunications \cite{chiang2007layering}, water and sewer systems \cite{prakash2005targeting}, and many others.
	These networked systems are prone to disruption by either natural causes, such as extreme weather events and aging equipment, or purposeful attack, such as terrorism \cite{albert2000error, crucitti2004error}.
	In power grid systems, small local failures have been known to cascade to blackouts affecting large swaths of a state or a country \cite{albert2004structural}.
	In a road system, small incidents can lead to large scale congestion \cite{bell2008attacker}.
	Attacks on networked systems can be either structural, where links in the underlying graph are damaged or destroyed, or dynamical, where a disruptive term is added to the dynamical equations that govern the behavior of the system.
	While our approach can be extended to encompass both structural and dynamical attacks, for the sake of simplicity, here we focus on dynamical attacks.
	Some examples of dynamical attacks on networked systems are pollutants introduced in a hydraulic network \cite{kessler1998detecting} or the spreading of viruses in networked computers \cite{pastor2001epidemic}.\\
	\indent \\We examine the behavior of simple dynamical networks, when they are attacked by one or more external signals which perturb the dynamics of the network nodes.
	To illustrate this situation, a ten node network where three of the nodes are under attack is shown in Fig.~\ref{fig:network1}.
	We assume that the networks at hand have an ability to defend themselves against attacks with appropriate counter-measures, which we model as actuators located at a subset of the nodes in the network.
	In Fig.~\ref{fig:network1}(e), nodes 1, 3, and 10 are attached to actuators, and so these nodes we define as \textit{defenders} (equivalently \textit{driver nodes} as they are defined in much of the complex network literature \cite{liu2011controllability}).
	Defenders can take many different forms in the networked systems described above, such as traffic signals and GPS routing in road networks, purposely tripping lines in a power grid in case of shedding, or quarantining a portion of a computer network when attacked by viruses.\\
	\\
	As a reference example, in this paper we consider a power grid, with nodes representing buses and edges representing transmission lines \cite{amini2016dynamic}. Both generation and a load can be present at each bus. We assume that some of the loads are vulnerable to attacks, in which case ancillary generation at the other nodes can be used to mitigate the effects of the attack. This problem is discussed in more detail in what follows.\\
	\\
	\clearpage
	
	Here, we consider a network which has stabilizing self-loops so that the adjacency matrix is Hurwitz. This ensures that after any perturbation of the states away from the origin, the states will return to the origin. Next, we add external attackers to the system attached to nodes 2, 5, and 7. In Fig.~\ref{fig:network1}, we color nodes 2, 5, and 7 cyan as they are the attacked nodes. The time evolution of the states of those nodes directly attacked, and any nodes downstream such as nodes 3, 6, and 10 are now diverging. On the other hand, any node upstream of the attackers are not effected by the attack and so they will converge to the origin. To counter the attacks, we add external control inputs attached to the red nodes 1, 3, and 10, which we call defender nodes. Thanks to the control action exerted by the defender nodes, the attack can be mitigated and now all nodes return to the origin again.
	
	For simplicity, the network dynamics is described by a linear model,
	\begin{equation}\label{eq:1}
	\dot{\textbf{x}}(t)=A{\textbf{x}}(t)+H{\textbf{w}}(t)+B{\textbf{u}}(t)
	\end{equation}
	where $\textbf{x}(t)=[x_1(t),..,x_n(t)]$ is the $n \times 1$ time-varying state vector, $\textbf{u}(t)=[u_1(t),..,u_m(t)]$ is the $m \times 1$ time-varying control input vector and $\textbf{w}(t)=[w_1(t),..,w_q(t)]$ is the $q \times 1$ time-varying vector representing the attackers. Hereafter, we look at the network using the fixed-end point minimum energy control problem for a system described by the linear dynamics shown in Eq. (1). Here $A=\{a_{ij}\}$ is a square $n \times n$ real adjacency matrix which has non-zero elements $a_{ij}$ if node $i$ receives a signal from node $j$ and is $0$ otherwise. The $n \times m $ matrix \emph{B} is the control input matrix and describes how the control inputs are connected to the nodes, i.e. the location of the defenders, namely $B_{ij}$ is different from zero if the control input $j$ is attached to node $i$ and is zero otherwise. The matrix \emph{H} models how the attackers affect the network nodes, namely $H_{ij}$ is different from zero if attacker $j$ is active on node $i$ and is zero otherwise. The matrix \emph{A} is Hurwitz and therefore by setting \textbf{w} = \textbf{0}  and \textbf{u}=\textbf{0}, the system asymptotically approaches the origin of state space, which represents the nominal healthy condition for the system.

	As explained in Sec.~\ref{Application}, the dynamics of a power grid can be cast in the form of Eq. (\ref{eq:1}),
	\begin{equation}\label{eq:2}
	\begin{bmatrix}
	\dot{\bm{\delta}} \\
	\dot{\bm{\theta}} \\
	\dot{\bm{\omega}}\\
	\end{bmatrix} =\textit{A}\begin{bmatrix}
	\bm{\delta} \\
	\bm{\theta} \\
	\bm{\omega }\\
	\end{bmatrix} + \textit{H} \begin{bmatrix} 
	0\\
	\bm{P}^{L}\\
	0\\ \end{bmatrix} + \textit{B} \begin{bmatrix} 
	0\\
	0\\
	\bm{P}^{M'}\\ \end{bmatrix}.
	\end{equation}
	where, the vector $\bm{\delta}$ = $[ \delta_{1},...,\delta_{n},]$ contains information on the voltage phase angles at generator buses, the vector $\bm{\theta}$=$[ \theta_{1},...,\theta_{n}]$ describes the voltage phase angles at load buses and the vector $\bm{\omega}$ = $[ \omega_{1},...,\omega_{n},]$ represents the frequency deviation at generator buses. The vector \textbf{\textit{P\textsuperscript{$L$}}}=$[P_{1}^{L},...,P_{n}^{L}]$  contains information on the power consumption at the load buses and the vector \textbf{\textit{P\textsuperscript{$M'$}}}=$[P_{1}^{M'},...,P_{n}^{M'}]$ represents the ancillary power generation (for more details see  section ~\ref{Application}.)\\
	\\
	We now introduce the strong assumption that a known model for the attackers exists. This assumption could be more or less unrealistic depending on the application to which we are applying this methodology; however, our results are general as they can be applied to a variety of models for the attackers' strategy and as we will see, they are to some extent independent of the attackers' specific strategy. The type of attack strategies we consider is either one of the following functions: (i) constant, (ii) linearly increasing, or (iii) exponentially increasing, which can be modeled as:\begin{equation}\label{eq:3}
	\dot{w}_i=s_iw_i+r_i \\
	\end{equation}
	where, $s_i$ and $r_i$ are constants. We consider the following three cases:\\
	(i) if $s_i=0$ and $r_i=0$ then attack strategy is constant.\\
	(ii) if $s_i=0$ and $r_i>0$ then attack strategy is linearly increasing.\\
	(iii) if $s_i>0$ and $r_i=0$ then attack strategy is exponentially increasing.\\
	Then, by incorporating the model for the attackers' behavior, we can rewrite Eq. (\ref{eq:1}) as follows
	\begin{equation}\label{eq:4}
	\dot{\tilde{\textbf{x}}}(t)=\tilde{A}{\tilde{\textbf{x}}}(t) + \mathbf{\tilde{r}} + \tilde{B}{\textbf{u}}(t)
	\end{equation}
	\begin{equation}\label{eq:5}
	{\textbf{y}}(t)=C\tilde{\textbf{x}}(t)
	\end{equation}
	where, 
	\begin{equation}\label{eq:6}
	\tilde{A} = \begin{bmatrix}
	A & \vdots & H\\
	\cdots & \cdots & \cdots\\
	0 & \vdots & S\\
	\end{bmatrix}, 
	\tilde{B} = \begin{bmatrix}
	B \\ \cdots \\0
	\end{bmatrix}, 
	C = \begin{bmatrix}
	I_n & \vdots & 0
	\end{bmatrix}, \mbox{and }
	\mathbf{\tilde{r}} = \begin{bmatrix}
	\textbf{0} \\ \cdots \\\textbf{r}
	\end{bmatrix}
	,    	
	\end{equation}
	\\

	Here  $\tilde{\textbf{x}}=[\textbf{x}^T, \textbf{w}^T]^T$ is the $n+q$ vector containing the states of the network nodes and attackers, the behavior of which is assumed to be known, $\textbf{y}(t)=[y_1(t),..,y_p(t)]$ is the $p \times 1$ time-varying vector of outputs, $S$=diag\{$s_1,..,s_q\}$ is the diagonal matrix that contains information on the attackers strengths and the vector $\mathbf{r}$=[$r_1,...,r_q$] describes the attackers strategies (see Eq. (3)). The matrix $\tilde{\emph{A}}$  now has a block triangular structure and is non-Hurwitz, due to the attackers' dynamics. The matrix \emph{C} relates the outputs $\textbf{y}(t)$ to the state $\tilde{\textbf{x}}(t)$. In this particular case, \textbf{y}(t) coincides with $\textbf{x}(t)$ in Eq. (\ref{eq:1}), i.e., it selects the states of the nodes but not those of the attackers.
	
	When $\textbf{u}=\textbf{0}$, the time evolution of the network nodes deviates from the origin due to the influence of the attackers. The question we will try to address is the following: how can we design an optimal control input that in presence of an attack, will set the state $\textbf{x}(t_{f})=\textbf{0}$ at some preassigned time $t_{f}$. The time $t_{f}$ can be thought as the required time to neutralize the attackers, so that for $t> t_{f}$, the network has returned to its healthy state and the control action is no needed anymore.  Here, without loss of generality, we assume the optimal control input to be the one that minimizes the energy function,
	\begin{equation}\label{eq:7}
	E=\int_{t_0}^{t_f} \textbf{u}^{T}(t)\textbf{u}(t) dt
	\end{equation}

	The control input $\textbf{u}^*(t)$ that satisfies the constraints and minimizes the control energy is equal to \cite{klickstein2017energy}:
	\begin{equation}\label{eq:8}
	\textbf{u}^*(t)=B^Te^{\tilde{A}^T(t_f-t_0)}C^T(CWC^T)^{-1} \times [\textbf{y}_f-Ce^{\tilde{A}^T(t_f-t_0)}\textbf{x}_0-CF(t_f)\mathbf{\tilde{r}}]
	\end{equation}
	where, $F(t_f) = \int_{t_0}^{t_f} e^{\tilde{A}(t_f - \tau)} d\tau$. The corresponding optimal energy is $\emph{E}^*$=$\int_{t_0}^{t_f} \textbf{u}^{*T}(t)\textbf{u}^{*}(t) dt$. First, we define the controllability Gramian as a real, symmetric, semi-positive definite matrix
	\begin{equation}\label{eq:9}
	W=\int_{t_{0}}^{t_{f}} e^{\tilde{A}({t_{f}-\tau})} {\tilde{B}}{\tilde{B}}^T e^{{\tilde{A}^T}({t_{f}-\tau})} dt
	\end{equation}
	Following \cite{klickstein2017energy},  the minimum control energy can be computed and is equal to
	\begin{equation}\label{eq:10}
	\begin{aligned}
	E^*
	& =(\mathbf{y}_f-Ce^{\tilde{A}({t_{f}-\tau})}-CF(t_f)\mathbf{\tilde{r}})^T (CWC^T)^{-1} (\mathbf{y}_f-Ce^{\tilde{A}({t_{f}-\tau})}-CF(t_f)\mathbf{\tilde{r}}) dt \\
	& =\bm{\beta}^T W^{-1}_{p} \bm{\beta} \\
	\end{aligned}
	\end{equation}
	where the vector $ \bm{\beta} = Ce^{\tilde{A}(t_f-t_0)}\textbf{x}_0 +CF(t_f)\mathbf{\tilde{r}}- \textbf{y}_f $ is the control maneuver and $W_{p}=CWC^T$ is the $p \times p$ symmetric, real, non-negative definite output controllability Gramian. The smallest eigenvalue of the output controllability Gramian, $\mu_{1}$, is nonzero if the system is output controllable. If this condition is satisfied, $\mu_{1}$ usually dominates the expression for the minimum control energy \cite{klickstein2017energy}. 
	\subsection{Effect of the Attackers on Output Controllability Gramian}	
	
	Consider the system with attackers (\ref{eq:4}) and (\ref{eq:5}). 
	We write,
	\begin{equation*}
	\begin{aligned}
	e^{\tilde{A}t} & = \begin{bmatrix}
	e^{At} & \vdots & F_1(t) \\
	\cdots & \cdots & \cdots\\
	0& \vdots & e^{St}\\
	\end{bmatrix}, \mbox{ where } F_1(t) = \int_0^1 e^{(1-\tau)At} H t e^{\tau St} d\tau  \cite{dieci2000pade}.
	\\ \mbox{Moreover,}\\
	\end{aligned}	
	\end{equation*}
	\\[0.1in]
	$\tilde{B}\tilde{B}^T = \begin{bmatrix}
	B \\ \cdots \\0
	\end{bmatrix} \begin{bmatrix}
	B^T & \vdots & 0^T 
	\end{bmatrix} = \begin{bmatrix}
	BB^T& \vdots& 0 \\ \cdots & \cdots & \cdots \\0& \vdots & 0
	\end{bmatrix} $,	 \\
	
	The controllability Gramian,\begin{equation}\label{eq:11}
	\begin{aligned}
	W &  = \int_{t_0}^{t_f} e^{\tilde{A}t}\tilde{B}\tilde{B}^Te^{\tilde{A}^Tt} dt\\
	&  = \int_{t_0}^{t_f} \begin{bmatrix}
	e^{At} & \vdots & F_1(t)\\
	\cdots & \cdots & \cdots\\
	0& \vdots & e^{St}\\
	\end{bmatrix} \begin{bmatrix}
	BB^T& \vdots& 0 \\ \cdots & \cdots & \cdots \\0& \vdots & 0
	\end{bmatrix}  \begin{bmatrix}
	e^{A^Tt} & \vdots & 0 \\
	\cdots & \cdots & \cdots\\
	F_1^T(t)& \vdots & e^{S^Tt}\\
	\end{bmatrix} dt\\
	&  = \int_{t_0}^{t_f} \begin{bmatrix}
	e^{At}BB^T  e^{A^Tt}& \vdots& 0 \\ \cdots & \cdots & \cdots \\0& \vdots & 0
	\end{bmatrix}    dt\\
	& = \begin{bmatrix}
	W_p & \vdots& 0 \\ \cdots & \cdots & \cdots \\0& \vdots & 0
	\end{bmatrix} 
	\linebreak 
	\end{aligned}
	\end{equation}
	
	Note that $W_p \in \mathbb{R}^{n\times n}$ does not depend on on the matrices $S$ and $E$ i.e., it is independent of the location of the attackers and the strength of the attackers	.
	The output controllability Gramian,
	\begin{equation}\label{eq:12}
	\begin{aligned}
	C W C^T & = \begin{bmatrix}
	I & \vdots & 0
	\end{bmatrix} \begin{bmatrix}
	W_p & \vdots& 0 \\ \cdots & \cdots & \cdots \\0& \vdots & 0
	\end{bmatrix} \begin{bmatrix}
	I \\ \cdots \\ 0
	\end{bmatrix}=W_p.\\
	\end{aligned}
	\end{equation}
	\linebreak
	If the pair $(A,B)$ is controllable, the matrix $W_p$ is positive definite and thus invertible
	\cite{murota1990note}\cite{rugh1996linear}.	
	\subsection{Effect of the Attackers on Control Maneuver}
	We have already defined the control maneuver as
	\begin{equation}\label{eq:13}
	\begin{aligned}
	\bm{\beta}& = Ce^{\tilde{A}(t_f-t_0)}\textbf{x}_0 +CF(t_f)\mathbf{\tilde{r}}- \textbf{y}_f
	\end{aligned}
	\end{equation}
	According to our assumptions we set $\textbf{y}_f = \textbf{0}$ (target state coincides with the origin).
	We write the eigenvalue equation for the matrix $\tilde{A}$, $\tilde{A}=V \Lambda  V^{-1}$, where the eigenvector matrix  $V = \left[\textbf{v}_1 \quad \vline \quad \textbf{v}_2 \quad \vline \quad  \cdots \quad \vline \quad \textbf{v}_{n+q} \right] $ and the eigenvalue matrix $\Lambda = diag\{\lambda_1,\cdots,\lambda_q,\lambda_{q+1},\cdots,\lambda_{q+n}\}$\\     
	where,  $\lambda_1\ge \cdots \ge \lambda_q\ge \lambda_{q+1}\ge \cdots\ge \lambda_{q+n}$. Note that because of the block diagonal structure of $\tilde{A}$ and the assumption that the matrix $A$ is Hurwitz the first $q$  eigenvalues of $\tilde{A}$, which correspond to the attackers dynamics,  $\lambda_{i}=s_{i}$, $i=1,...,q$.

	We write, $\mathbf{x_{0}}=  \sum_i c_i \textbf{v}_i = V \textbf{c}$ , where the vector $\textbf{c} = 
	\left[ c_1,c_2,\cdots,c_{n+q}\right]
	$,\\ 
	
	Now from Eq. (13),
	\begin{equation}\label{eq:14}
	\begin{aligned}
	\bm{\beta}
	& = Ce^{\tilde{A}(t_{f}-t_{0})}\textbf{x}_0 =  C(\sum\limits_{i=1}^{n+q} c_ie^{\lambda_{i}(t_{f}-t_{0})}\textbf{v}_i+\sum\limits_{i=1}^{n+q} g_iJ_i\textbf{v}_i )\\
	\end{aligned}
	\end{equation} 
	where, $J_i$=$\frac{e^{\lambda_{i}(t_{f}-t_{0})}-1}{\lambda_{i}}$. For large $t_f$ the above equation can be approximated as
	\begin{equation} \label{approx}
	\bm{\beta}\approx  C\sum\limits_{i=1}^{q} \Big( c_ie^{s_{i}(t_{f}-t_{0})}+g_i\frac{e^{s_{i}(t_{f}-t_{0})}-1}{s_{i}}\Big) \mathbf{v}_i
	\end{equation}
	
	We write,
	\begin{equation}\label{eq:15}
	\begin{aligned}
	\\\bm{\beta}
	& = {\beta}\bm{n}\\
	\end{aligned}
	\end{equation}  
	where $\bm{n}$ is the vector with norm equal to 1 having the same direction as $\bm{\beta}$. We see from Eq.\ \eqref{approx} that for large $t_{f}$, the order of magnitude of $\bm{\beta}$ is determined by the number and strengths of attackers (i.e., $s_i$, $i=1,..,q$).   
	\\
	We now express the symmetric matrix $W_{p}$ in terms of its eigenvalues and eigenvectors. $W_{p}\bm{w}_{i}=\mu_{i}\bm{w}_{i},\mbox{ where } i=1,...,N :
	W_{p}^{-1}=\sum\limits_{i=1}^n \mu_{i}^{-1}\bm{w}_{i}\bm{w}_{i}^T.\\$
	
	Replacing $W_{p}^{-1}$ into the eq. (\ref{eq:10})
	\begin{equation}\label{eq:16}
	\begin{aligned}
	E^*
	&=\bm{\beta^T}\sum\limits_{i=1}^n \mu_{i}^{-1}\bm{w}_{i}\bm{w}_{i}^T\bm{\beta}, \mbox{ where } \mu_1\le\mu_2\le.....\le\mu_N\\
	&\approx \beta^2\sum\limits_{i=1}^n (\bm{n}^T\bm{w}_{i})^2\mu_{i}^{-1}\\
	\end{aligned}
	\end{equation}
	where the approximation holds, when $\mu_{1} \ll \mu_{2}$ and when $ \bm{n}^T \bm{w}_1\ne0$. Thus we can write
	\begin{equation}\label{eq:17}
	\begin{aligned}
	E^*
	&\approx \beta^2\mu_{1}^{-1}(\bm{n}^T \bm{w}_1)^2\\ 
	&=E_1E_2E_3,\\
	\end{aligned}
	\end{equation}
	where $E_1={\beta}^2$ corresponds to the strength of the attackers, $E_2=\mu_1^{-1}$ does not depend on the attackers but depends on the network topology and the location of the defenders and $E_3=(\bm{n}^T \bm{w}_1)^2$ depends on the distribution of attackers and defenders over the network. Note that the vector $\bm{w}_1$ is the eigenvector of $W_p$ associated with its smallest eigenvalue. The term $\bm{n}^T\bm{w}_1$ measures the angle between two vectors both having norm 1, thus $0\le$$(\bm{n}^T \bm{w}_1)^2$$\le1$. 
	
	Now consider a simple ten node network in Fig. 1 and  place the defenders on three nodes (nodes colored red in Fig. 1(e)). We have considered the effect of different choices of the attackers as can be seen from Fig. 2. The smallest eigenvalue of the output controllability Gramian $W_{p}$ remained constant as the number and position of the attackers was varied.
	
	\begin{figure}[H]
		\centering
		\begin{tabular}{c}
			\includegraphics{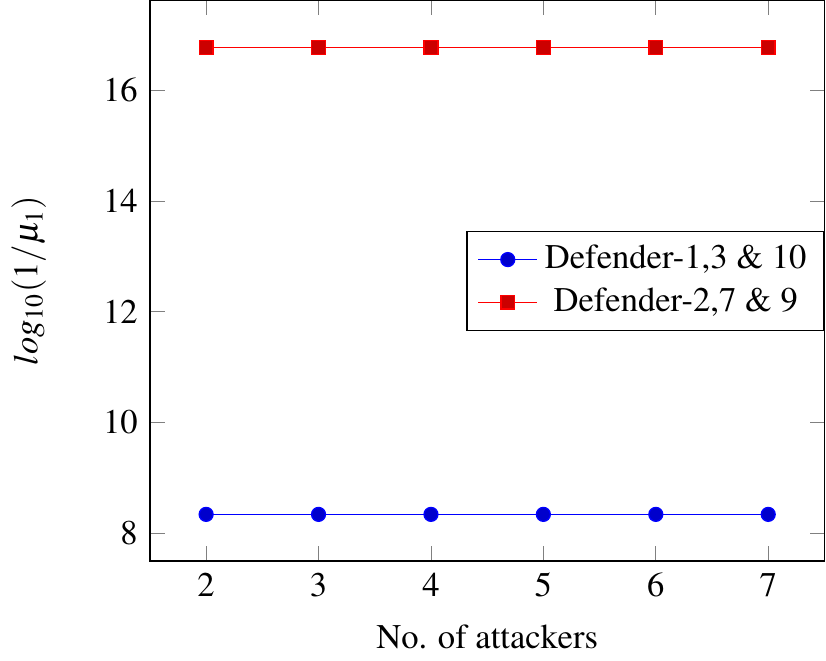}
		\end{tabular}
		\caption{log10(1/$\mu_{1}$) for the network shown in Fig.~\ref{fig:network1} as we increased the number of nodes subject to attacks. (\protect\markerone)  symbols: Defenders are placed on nodes 1, 3 \& 10 ; (\protect\markerfour) symbols: Defenders are placed on nodes 2, 7 and 9. Attackers are chosen in a random order, but ensuring that no node is both an attacker and a defender.}
		\label{fig:(2)}
	\end{figure}
	
	Figure~\ref{fig:(2)} also illustrates the case that the same network is subjected to attack changing only the position of the defenders, now at nodes 2, 7, and 9. Again we see that the minimum eigenvalue of the Gramian ($\mu_1$) is independent of the number and position of the attackers. However, we see that the $\mu_1$ depends on the location of the defenders.

	
	We have come to the initial conclusion that we can determine for different networks, and different locations of attackers and defenders, the minimum control energy needed to control a network under attack in a preassigned time. Our main result is that the expression for the minimum control energy can be approximated as follows:
	$\emph{E}^*\approx E_{1} E_{2} E_{3}$.
	While $E_{1}$ depends on the position of the attackers but not on the network topology, $E_{2}$ depends on the matrices $A$ and $B$ (on the Gramian), but not on the number, position and strength of the attackers, and the quantity $E_{3}$ depends on the distribution of attackers and defenders over the network. This is investigated in more detail in the following sections.
	
	\section{Analysis of network topologies}
	In this section we investigate how the control energy varies as we vary the position of attacked nodes and defenders over several networks. In all the simulations that follow, we set  $A_{ij}=A_{ji}=1$ if a connection exists between node $i$ and $j$ and $A_{ij}=A_{ji}=0$ otherwise. We also set the matrices $B$ and $H$ to be composed of different versors as columns, which indicates each attacker and/or defender is localized at a given node (in particular each attacker is attached to one and only one attacked node).
	In this section, in order to compute the quantities $\bm{n}^T\bm{w}_1$ and $\beta^2$, we add a small noise term to the entries on the main diagonal of the adjacency matrix $A$, $A_{ii} \leftarrow A_{ii} + \phi_i$, $i=1,...,N,$ where $\phi_{i}$ is a random number uniformly chosen in the interval $\in [0, \epsilon]$. This is done to ensure the pair $(A,B)$ is controllable, see e.g., \cite{yan2015spectrum,klickstein2017energy}. 
	
	\subsection{Chain networks}
\begin{figure}[H]
		\centering
		\begin{tabular}{c}
			\includegraphics{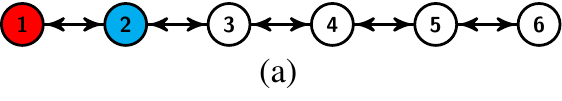}\\
			\begin{tabular}{cc}
				\includegraphics{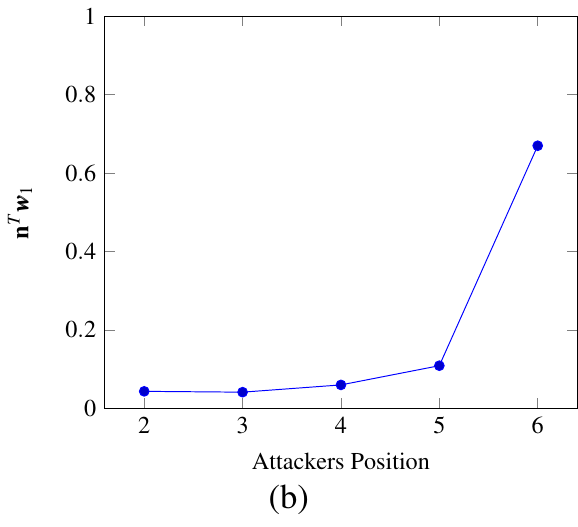}&
				\includegraphics{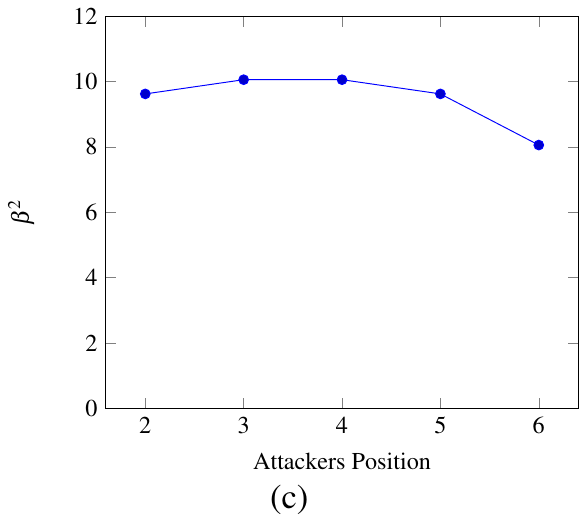}\\
			\end{tabular}
		\end{tabular}
		\caption{(a) Bidirectional chain network. The defender node is in red and the attacked node is in cyan. (b) $\mathbf{n}^T \mathbf{w}_1$ versus the position of the attacker. (c) Variation of $\mathbf{\beta^2}$  as the position of the attacker is varied. We perform calculations setting $t_{f}$=1, $s_i = 2.5$, $r_i = 0$ and $\epsilon=10^{-2}$.}
		\label{fig:line6}
	\end{figure}

	Now we investigate how $E_{1}$ and $E_{3}$ vary in the six node bidirectional chain network shown in Fig.~\ref{fig:line6}(a) We keep the position of the defender fixed at node 1 as indicated in  figure 3. Then we vary the position of the attacked nodes over the chain.

	We see that the term $\bm{n}^T\bm{w}_1$, corresponding to $E_{3}$, generally increases  when we increase the distance between the defender node and the attacked node. The term $\bm{n}^T\bm{w}_1$ is largest when the attacker is at node 6, i.e the node which is  farthest from the defender. Also, we see a small variation in the terms of $\beta^2$ as we change the position of the attacker as above. However, the effect of varying the position of the attacker on $\beta^2$ is less pronounced than on $\bm{n}^T\bm{w}_1$.
	Overall, these results are consistent with previous studies on target control of networks where the control energy was found to increase with the distance between driver nodes and target nodes \cite{klickstein2018energy}.
	\begin{figure}[H]
		\centering
		\begin{tabular}{c}
			\includegraphics{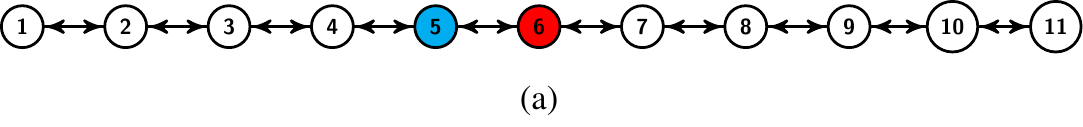}\\
			\begin{tabular}{cc}
				\includegraphics{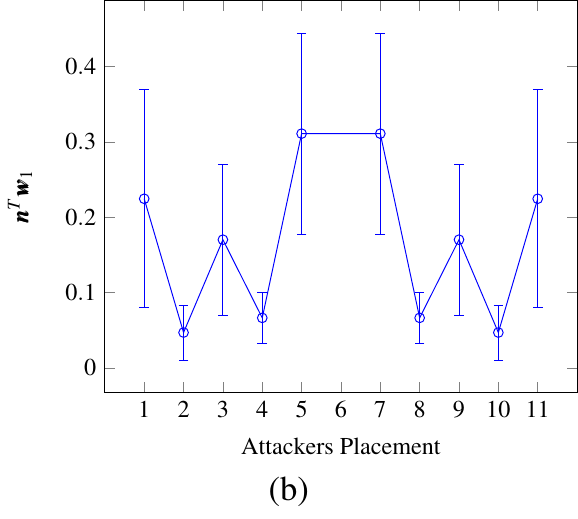}&
				\includegraphics{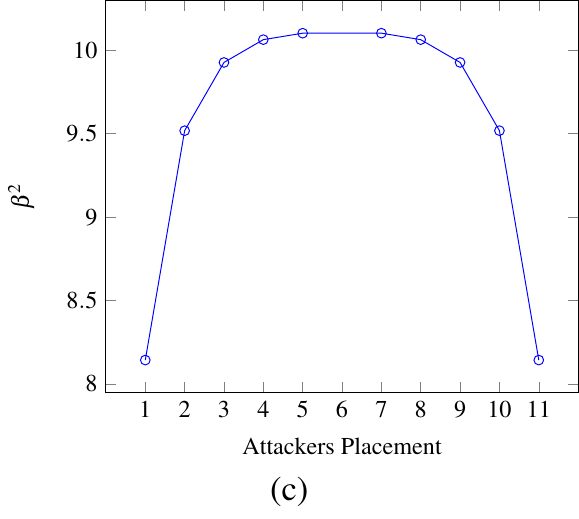}\\
			\end{tabular}
		\end{tabular}
		\caption{(a) A chain network with defender at the center node. (b) Plot of  $\bm{n}^T\bm{w}_1$ vs. the position of the attacked node. (c) Variation of $\beta^2$ vs. the position of the attacked node. We perform calculations setting $t_{f}$=1, $s_i = 2.5$, $r_i = 0$ and $\epsilon=10^{-2}$. The bars represent the standard deviation taken over 100 different realizations.}
		\label{fig:line11}
	\end{figure}
	
	Figure 4 shows the case that the defender is placed at the center node of the chain network. Here we see that the quantity $\beta^2$  decreases as the distance from the defender node and the attacked node increases. However, the quantity ${\mathbf{n}}^T {\mathbf{w}}_1$ displays a much more complex and somehow surprising behavior, also distinctly different from that observed in Fig.\ 3. Namely, we see that the quantity ${\mathbf{n}}^T {\mathbf{w}}_1$ alternatively increases and decreases as the position of the attacker is moved over the chain. This type of behavior is different from what seen in the case of target control of networks\cite{klickstein2018energy}.
	
	\subsection{Star network}
	\begin{figure}[H]
		\centering
			\begin{tabular}{ccc}
				\includegraphics{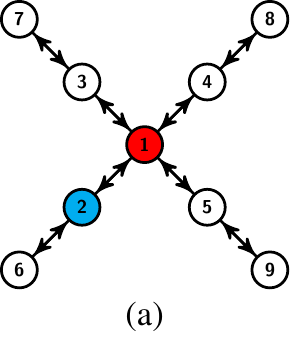}&
				\includegraphics{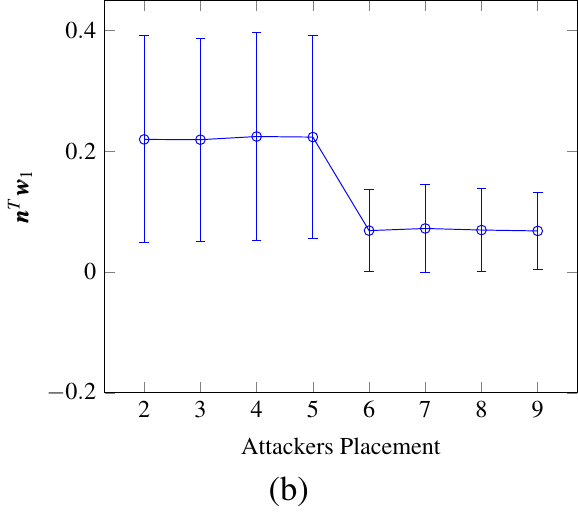}&
				\includegraphics{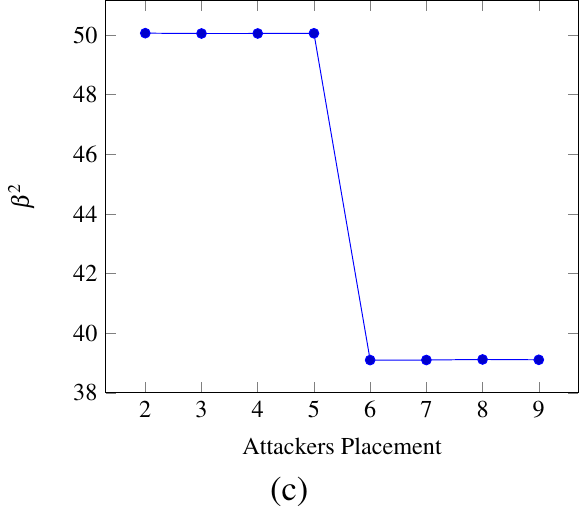}\\
		\end{tabular}
		\caption{(a) A star network. (b) Plot of  $\bm{n}^T\bm{w}_1$ vs position of the attacker. (c) $\beta^2$ vs. the position of the attacked nodes. We perform calculations setting $t_{f}$=1, $s_i = 2.5$, $r_i = 0$ and $\epsilon=10^{-2}$. The bars represent the standard deviation taken over 100 different realizations.}
		\label{fig:star}
	\end{figure}
	
	We now consider the case of the star network in Fig.~\ref{fig:star}(a) with defender at node 1 and the position of the attacked node varied  from node 2 to 9. We see that the value of $\bm{n}^T\bm{w}_1$ when the position of the attacked node is in the first layer of the star network (i.e on nodes 2, 3, 4 and 5) is nearly constant over that layer. When the attacker is on the second layer (i.e on nodes 6, 7, 8 and 9) the value of $\bm{n}^T\bm{w}_1$ is also nearly  constant. In Fig.~\ref{fig:star}(b) we see a similar pattern for $\beta^2$ as we saw for $\bm{n}^T\bm{w}_1$ in Fig.~\ref{fig:star}(b). The value of $\beta^2$ for the first layer is equal and so is for the second layer. However, when comparing the two layers, we see from both panels (b) and (c) in Fig.~\ref{fig:star} that surprisingly the energy to control the star network  decreases with the distance between the attacked node and the defender over the  network.
	\pagebreak
	\subsection{Ring network}
\begin{figure}[H]
		\centering
		\begin{tabular}{cc}
			\includegraphics{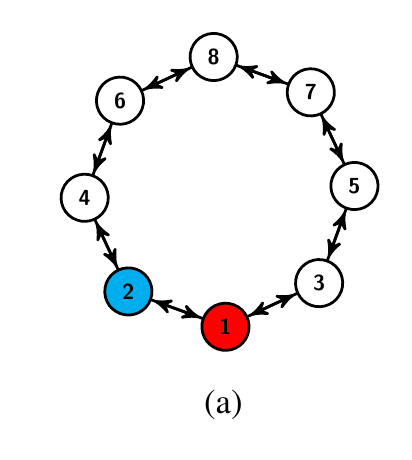}&
			\includegraphics{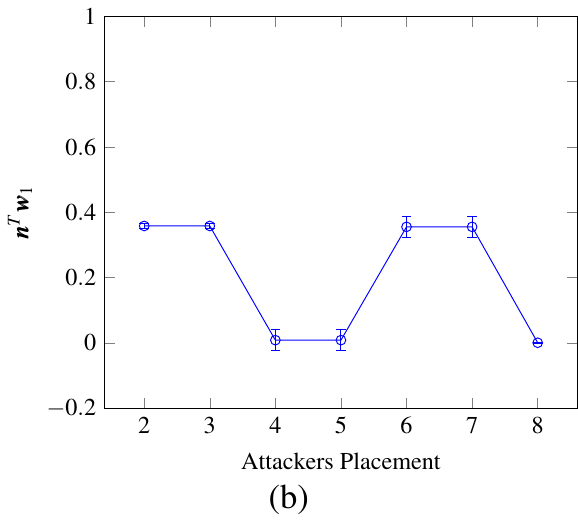}\\

			\includegraphics{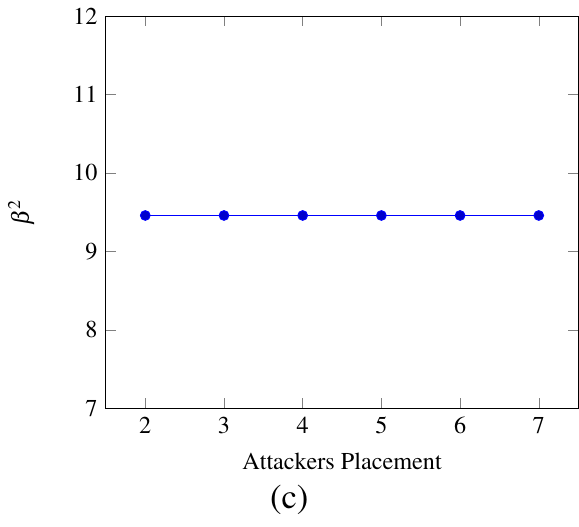}&
			\includegraphics{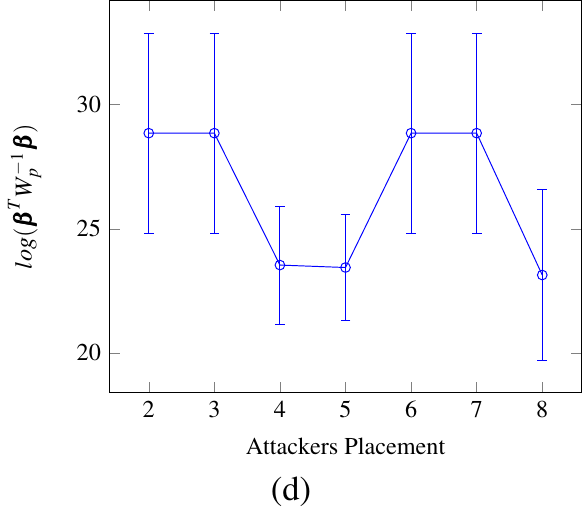}\\
	
		\end{tabular}
		\caption{(a) An eight node ring network with defender at node 1. (b) Plot of $\bm{n}^T\bm{w}_1$ vs the position of the attacked node. (c) Plot of  $\beta^2$ vs the position of the attacked node. (d) Total energy $E^*$ as the position of the attacked node is varied. The bars represent standard deviations over 100 different realizations. We perform calculations setting the final time $t_{f}$=1, $s_i = 2.5$, $r_i = 0$ and $\epsilon=10^{-2}$.}
		\label{fig:ring}
	\end{figure}

	We now consider a small ring network of 8 nodes (shown in Fig.~\ref{fig:ring}(a)). The defender is at node 1 and the attacker can be at any other node.
	From Fig.~\ref{fig:ring}(b) we see that the value of  $\bm{n}^T\bm{w}_1$ varies with the distance between the attacked and defender nodes over the ring (nodes 2,3 at distance 1, nodes 4,5 at distance 2, nodes 6,7 at distance 3, and node 8 at distance 4). However, the variation is, once again, non monotonous with respect to the distance. In particular, we do not see that the energy to control the attack monotonically increases with the distance between attacker and defender. Fig.~\ref{fig:ring}(c) shows that in this case $\beta^2$ is independent of the position of the attacker over the ring network. Fig.~\ref{fig:ring}(d) shows the total energy $E^*$ from Eq. (\ref{eq:10}), which is consistent with Fig.~\ref{fig:ring}(b).
	
	\subsection{Scale free networks}
	\begin{figure}[H]
		\centering
		\begin{tabular}{cc}
			\includegraphics{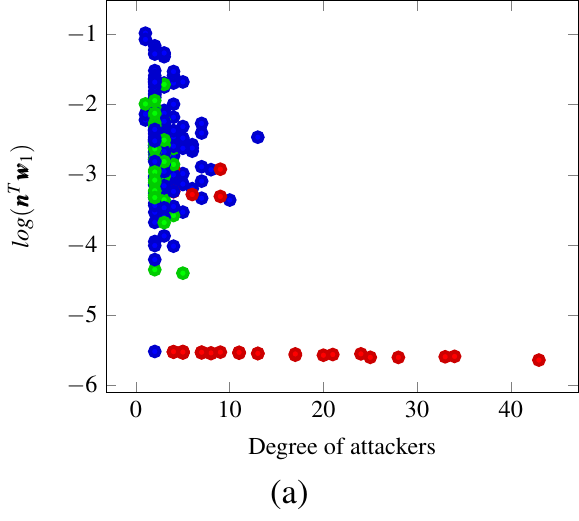}&
			\includegraphics{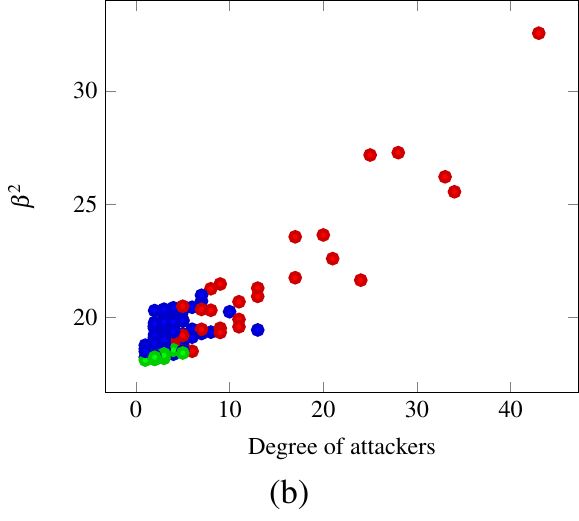}\\
			
		\end{tabular}
		\caption{(a) Plot of  $\bm{n}^T\bm{w}_1$ vs the degree of the attacked node for a 300 scale free network with average degree 2. (b) Plot of  $\beta^2$ vs the degree of the attacked node. (\protect\markerthree)   symbol indicates attacked nodes with $\Delta=3$. (\protect\markerone)   symbol indicates attacked nodes with $\Delta=2$. (\protect\markertwo)   symbol indicates attacked nodes with $\Delta=1$. We perform calculations setting the final time $t_{f}$=1 with $s_i = 2.5$,  $r_i = 0$ and $\epsilon =10^{-2}$.
		}
		\label{fig:scalefree}
	\end{figure}
	Here we consider a 300 node Barabasi Albert scale free network \cite{liu2011controllability} with average degree 2. We select $10 \%$ of the nodes to be defenders, and position them so to ensure that the pair $(A,B)$ is controllable \cite{klickstein2017energy}. We then vary the choice of a single attacked node over the network, one by one,  excluding the defender nodes. For each selection, we compute the minimum shortest distance $\Delta$ between the attacked node and the defender nodes, 
	\begin{equation}
	\Delta=\min\limits_{d} \mbox{shortest distance}(a,d),
	\end{equation}
	where $a$ indicates the attacked node and $d$ the defender nodes. 
	Each point in Fig.~\ref{fig:scalefree}(a) indicates the value of 
	$\bm{n}^T\bm{w}_1$ for a given choice of the attacked node versus the degree of the attacker. As can be seen, the quantity $\bm{n}^T\bm{w}_1$ varies over several order of magnitude for different choices of the attacked nodes. In particular certain nodes are \emph{weak attackers} as the required control energy is particularly low when these nodes are subject to an attack. Fig.~\ref{fig:scalefree}(b) indicates the value of $\beta^2$ for the given choice of attacker node versus the degree of the attacker. The quantity $\beta^2$ increases as the degree of the attacked nodes increases and the quantity $\Delta$ decreases. While attacked nodes with $\Delta=1$ tend to have a slightly higher value of $\beta^2$, the value of $\bm{n}^T \bm{w}_1$ is typically at least one order of magnitude lower, indicating that the minimum control energy $E$ is much lower for these nodes. Overall the figure shows that the degree of a node is not a good predictor for a weak attacker, as these are nodes of all possible degrees.  However, the parameter $\Delta$ appears to be a good indicator for a weak attacker, as these have typically $\Delta=1$, i.e., they are neighbors of at least one defender.

	\section{An example of application of the ANALYSIS TO INFRASTRUCTURE NETWORKS}  \label{Application}
	The analogy of networks with attackers is presented in \cite{amini2016dynamic}. Here, they use an IEEE 39 bus system where dynamic load altering attack is used to destabilize the system.\\
	
	The power system dynamics can be described as follows \cite{amini2016dynamic}:\\
	\begin{equation}\label{eq:18}
	\begin{bmatrix}
	I & 0 & 0 & 0\\
	0 & I & 0 & 0\\
	0 & 0 & -M & 0\\
	0 & 0 & 0 & 0\\
	\end{bmatrix}\begin{bmatrix}
	\dot{\bm{\delta}} \\
	\dot{\bm{\theta}} \\
	\dot{\bm{\omega}}\\
	\bm{\dot{\varphi}}\\
	\end{bmatrix} =\begin{bmatrix}
	0 & 0 & I & 0\\
	0 & 0 & 0 & -I\\
	K^{I}+H^{GG} & H^{GL} & K^{P}+D^{G} & 0\\
	H^{LG} & H^{LL} & 0 & D^{L}\\
	\end{bmatrix}\begin{bmatrix}
	\bm{\delta} \\
	\bm{\theta} \\
	\bm{\omega}\\
	\bm{\varphi}\\
	\end{bmatrix} + \begin{bmatrix}
	0\\
	0\\
	0\\
	I\\
	\end{bmatrix}\bm{P^{L}} 
	\end{equation}
	\\
	\\
	Equation (20) can be rewritten as follows, after setting $\dot{\bm{\varphi}}$ to zero and replacing in the equations for the time evolution of $\bm{\delta}$, $\bm{\theta}$, and $\bm{\omega}$, 
	
	\begin{equation}\label{eq:19}
	\begin{bmatrix}
	\dot{\bm{\delta}} \\
	\dot{\bm{\theta}} \\
	\dot{\bm{\omega}}\\
	\end{bmatrix} = \begin{bmatrix}
	I & 0 & 0 \\
	0 & D^{L^{-1}} & 0 \\
	0 & 0 & -M^{-1}  \\
	\end{bmatrix} \begin{bmatrix}
	0 & 0 & I \\
	H^{LG} & H^{LL} & 0\\
	K^{I}+H^{GG} & H^{GL} & K^{P}+D^{G} \\
	\end{bmatrix}\begin{bmatrix}
	\bm{\delta} \\
	\bm{\theta}\\
	\bm{\omega}\\
	\end{bmatrix} + \begin{bmatrix}
	0\\
	D^{L^{-1}}\\
	0\\
	\end{bmatrix}\bm{P^{L}}
	\end{equation}
	\\
	
	Let us assume we can add ancillary generator in our power grid system to compensate for over- and under-frequency disruptions. Then the mechanical power input $\bm{P_{i}^{M}}$ at the $i$ generator with ancillary generation power $\bm{P_{i}^{M'}}$ is given by \\
	\begin{equation}\label{eq:20}
	\bm{P_{i}^{M}}  = - (K_{i}^{P} \omega_i +  K_{i}^{P} \int_{0}^{t} \omega_i + \bm{P_{i}^{M'}}). 
	\end{equation}
	\\
	
	Now the total power grid system with load attack on $\textbf{P}^L$ and ancillary generation $\textbf{P}^{M}$ can be written in the form of Eq.\ (22) as follows
	\begin{equation}\label{eq:21}
	\begin{bmatrix}
	\dot{\bm{\delta}} \\
	\dot{\bm{\theta}} \\
	\dot{\bm{\omega}}\\
	
	\end{bmatrix}=\textit{A}\begin{bmatrix}
	\bm{\delta} \\
	\bm{\theta}\\
	\bm{\omega}\\ 
	\end{bmatrix} + \textit{H} \begin{bmatrix} 
	0\\
	\bm{P^{L}}\\
	0\\ \end{bmatrix} + \textit{B} \begin{bmatrix} 
	0\\
	0\\
	\bm{P^{M}}\\ \end{bmatrix}.
	\end{equation}
	
	where,\\
	\\
	A= $\begin{bmatrix}
	I & 0 & 0 \\
	0 & {(D^{L})}^{-1}& 0 \\
	0 & 0 & -M^{-1}\\
	
	\end{bmatrix}$ $\times$ $\begin{bmatrix}
	0 & 0 & I \\
	H^{LG} & H^{LL} & 0 \\
	K^{I}+H^{GG} & H^{GL} & K^{P}+D^{G}\\
	
	\end{bmatrix}$ \\
	
	and\\
	
	H=$\begin{bmatrix}
	0\\
	{(D^{L})}^{-1} \\
	0\\
	
	\end{bmatrix}$\\
	
	B=$\begin{bmatrix}
	0\\
	0\\
	-M^{-1}\\
	
	\end{bmatrix}$\\
	
	The matrix \textit{A} is the system matrix, the matrix \textit{E} determines the effect and position of the attackers and the matrix \textit{B} the effect and position of the defenders.\\
	\section{\label{sec:level5}Conclusions}
	
	In this paper we have studied an optimal control problem on networks, where a subset of the network nodes are attacked and the goal is to contrast the attack using available actuating capabilities at another subset of the network nodes. Compared with previous work on optimal control of network\cite{liu2011controllability,klickstein2017energy,yan2015spectrum,klickstein2018energy}, we consider a situation in which the control action is implemented, while another external dynamics is also taking place in the network.\\ 
	We envision this work to be relevant to critical infrastructure networks (such as power grids), which are susceptible to attacks.
	While our results assume knowledge of the attacker's strategy, which is often unavailable, our analysis can used to the design infrastructure networks that are resistant to attacks. This can be done by considering all the possible attacks that can affect the network and for each case, compute the optimal control solution.
	We have studied how the minimum control energy varies  as the position of the $attackers$ and $defenders$ is varied over different networks such as chain, star, ring and scale free networks. Our main result is that the expression for the minimum control
	energy can be approximated by the product of three different quantities $E_1$$E_2$$E_3$. While $E_1$ depends on the position of the
	attackers but not on the network topology, $E_2$ depends on the matrices $A$ and $B$ (on the Gramian), and $E_3$ depends on the position of both the attacked nodes and defender nodes over the network.\\
	In chain, star and ring networks, we see that for a single attacker and a single defender, often the minimum control energy is not an increasing function of the distance between the attacked node and the defender node. However, for a scale free network with multiple defenders and a single attacker, we see that a good predictor for the strength of the attack is provided by the quantity $\Delta$ (the minimum distance between the defender nodes and the attacked node).
	
	\subsection*{ACKNOWLEDGEMNT}
	This work was supported by the National Science Foundation
	though NSF Grant No. CMMI- 1400193, NSF Grant
	No. CRISP- 1541148, ONR Grant No. N00014-16-1-
	2637, and DTRA Grant No. HDTRA1-12-1-0020.
	

\end{document}